\documentclass[useAMS,usenatbib,usegraphicx]{mn2e}
\title[Variation of the Arches non-thermal emission]{Variation of the X-ray non-thermal emission \\ in the Arches cloud}
\author[M. Clavel et al.]{M. Clavel,$^{1, 2}$\thanks{E-mail:
maica.clavel@apc.univ-paris7.fr (MC)} S. Soldi,$^{1}$ R. Terrier,$^{1}$ V. Tatischeff,$^{3}$ G. Maurin,$^{4}$ G. Ponti,$^{5}$
\newauthor
A. Goldwurm$^{1, 2}$ and A. Decourchelle$^{6, 2}$\\
$^{1}$ AstroParticule et Cosmologie, Universit\'e Paris Diderot, CNRS/IN2P3, CEA/DSM, Observatoire de Paris, Sorbonne Paris Cit\'e; \\ \hspace{1.5mm} 10, rue Alice Domon et L\'eonie Duquet, 75205 Paris Cedex 13, France\\
$^{2}$ Service d'Astrophysique/IRFU/DSM, CEA Saclay; B\^at. 709, 91191 Gif-sur-Yvette Cedex, France\\
$^{3}$ Centre de Sciences Nucl\'eaires et de Sciences de la Mati\`ere, CNRS/IN2P3, Universit\'e Paris-Sud; 91405 Orsay, France\\
$^{4}$ Laboratoire d'Annecy-le-Vieux de Physique des Particules, Universit\'e de Savoie, CNRS/IN2P3; 74941 Annecy-le-Vieux, France\\
$^{5}$ Max-Planck-Institute for Extraterrestrial Physics, Garching, PSF 1312, 85741 Garching, Germany\\
$^{6}$ Laboratoire AIM (CEA/Irfu, CNRS/INSU, Universit\'e Paris VII), CEA Saclay; B\^at. 709, 91191 Gif-sur-Yvette Cedex, France}

\def\sgr{$\rm Sgr \, A^{\star}$}
\def\fe{Fe~K$\alpha$}

\def\nh{N$_{\rm H}$}
\def\n2h{N$_2$H$^+$}

\begin{document}

\date{Accepted 2014 June 19. Received 2014 June 19; in original form 2014 May 30}

\pagerange{\pageref{firstpage}--\pageref{lastpage}} \pubyear{2014}

\maketitle

\label{firstpage}

\begin{abstract}
The origin of the iron fluorescent line at 6.4\,keV from an extended region surrounding the Arches cluster is debated and the non-variability of this emission up to 2009 has favored the low-energy cosmic-ray origin over a possible irradiation by hard X-rays. By probing the variability of the Arches cloud non-thermal emission in the most recent years, including a deep observation in 2012, we intend to discriminate between the two competing scenarios. 
We perform a spectral fit of \textit{XMM-Newton} observations collected from 2000 to 2013 in order to build the Arches cloud lightcurve corresponding to both the neutral \fe\ line and the X-ray continuum emissions. 
We reveal a 30\% flux drop in 2012, detected with more than 4\,$\sigma$ significance for both components. This implies that a large fraction of the studied non-thermal emission is due to the reflection of an X-ray transient source.
\end{abstract}

\begin{keywords}
Galaxy: center -- X-ray: ISM -- ISM: clouds -- Cosmic rays
\end{keywords}

\section{Introduction}
The Arches cluster is a massive star cluster located within the Galactic center region, at about 11$'$ to the Galactic north-east of Sagittarius~A$^\star$. The X-ray emission of the cluster is associated with a thermal component that is thought to originate from collisions of winds from massive stars \citep{yusef-zadeh2002b,wang2006,tsujimoto2007}. A diffuse  and more extended non thermal emission including the neutral iron fluorescent line at 6.4 keV, has also been detected from a region directly surrounding the star cluster. 
This emission could be created either by the interaction of low-energy hadronic cosmic-rays with molecular material surrounding the cluster \citep{tatischeff2012} or by a strong X-ray irradiation as in other clouds of the central molecular zone \citep[CMZ;][and references therein]{clavel2013,ponti2013}.  
Both scenarios are compatible with the \textit{XMM-Newton} observations collected up to 2009 \citep{capelli2011b,tatischeff2012} and also with the latest \textit{NuSTAR} characterization of the higher-energy emission \citep{krivonos2014}. However, due to the constant \fe\ line emission and the absence of clear molecular counterpart of the X-ray emission, the low-energy cosmic-ray protons scenario has been favored \citep{tatischeff2012}.

In this paper, we add \textit{XMM-Newton} observations spanning three years more compared to the previous studies of the Arches non-thermal emission. In particular, we include data from the 2013 monitoring of \sgr\ and a deep observation obtained in 2012 during a scan of the CMZ. We find a significant decrease of both the 6.4 keV and the continuum emissions in the 2012 data set, suggesting that a significant part of the non-thermal emission is due to reflection. 
In Section~\ref{sec:dataAnalysis}, we present the data and its reduction. The results of our analysis are detailed in Section~\ref{sec:analysisResults}, and the origin of the non-thermal emission is discussed in Section~\ref{sec:discussion}.

\section{\textit{XMM-Newton} observations and data reduction}
\label{sec:dataAnalysis}
We analyzed all the \textit{XMM-Newton}/EPIC observations available since 2000 and including the Arches cluster region (reported in Tab.\,\ref{tab:obsID}). The data reduction was carried out using the \textit{XMM-Newton} Extended Source Analysis Software \citep[ESAS,][]{snowden2008} included in version 12.0.1 of the \textit{XMM-Newton} Science Analysis Software (SAS). Calibrated event lists were produced for each exposure using the SAS \textit{emchain} and \textit{epchain} scripts and ESAS mos-filter and pn-filter were used to exclude periods affected by soft proton flaring. 

\begin{figure*}
	\includegraphics[width=1.0\textwidth]{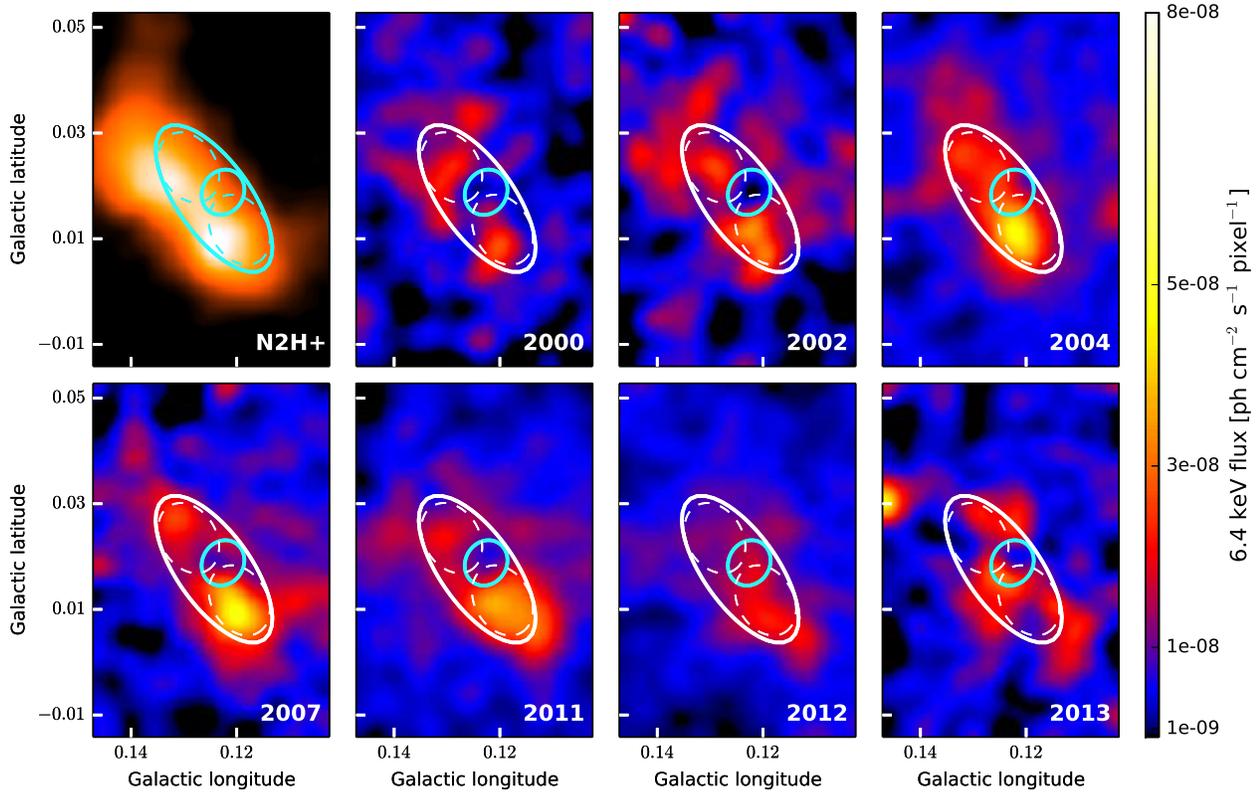}
	\caption{\textit{(Top Left)} Mopra \n2h\ map of the Arches cloud position integrated between --40 and --10\,km\,s$^{-1}$ \citep{jones2012}. A significant molecular contribution is there albeit a slight shift of the molecular emission to the south east compared to the X-ray emission of the Arches region. \textit{(Others)} Continuum and background subtracted \fe\ maps of the Arches cluster region for seven different years, from top left to bottom right: 2000, 2002, 2004, 2007, 2011, 2012 and 2013. The maps are displayed in Galactic coordinates and smoothed using a gaussian kernel of 20$''$ 
radius. The solid white ellipse is the cloud region, the cyan ellipse is the cluster region, and the two dashed ellipses are two cloud subregions: north and south. The overall cloud shows morphological variations from period to period with a clear decrease in the overall emission in 2012.
 }
	\label{fig:ArchesFeKa}
\end{figure*}

Background- and continuum-subtracted mosaic images have been created in the energy band 6.32--6.48~keV (Fig.\,\ref{fig:ArchesFeKa}). 
To produce them, we created the quiescent particle background (QPB) images, the count images and model exposure maps for each observation and each instrument, using \textit{mos-spectra},  \textit{pn-spectra}, \textit{mos\_back} and \textit{pn\_back} in the two energy bands 6.32--6.48~keV and 3--6~keV.
The combined exposure map was computed taking into account the different efficiencies of the three instruments. The contribution of the continuum to the 6.4~keV line was estimated using the extrapolation of the 3--6~keV emission and assuming an absorbed power-law spectrum with a photon index $\Gamma = 2$ and a column density $\rm N_{\rm H} = 7 \times 10^{22} \rm \, cm^{-2}$. 
The \textit{Chandra} analysis tool \textit{reproject\_image\_grid} was used to correctly reproject the maps of each instrument and each observation within the same year and to mosaic them. For each year, the total background mosaic and the estimated continuum mosaic were then subtracted from the 6.4~keV count mosaic and normalized by the total exposure to obtain the final count rate mosaics.

All spectra used in the analysis were extracted with the ESAS \textit{mos-spectra} and \textit{pn-spectra} scripts. 
In particular, the region defined to study the variability of the X-ray emission corresponds to the elliptical region where the brightest 6.4 keV emission is detected (largest ellipse in Fig.\,\ref{fig:ArchesFeKa}, referred as `Cloud' in Tab.\,\ref{tab:regions}). The Arches cluster (blue ellipse in Fig.\,\ref{fig:ArchesFeKa}, referred as `Cluster' in Tab.\,\ref{tab:regions}) has been excluded from the former region for the analysis. We point out that this spectral extraction region is very similar to the one used in \cite{tatischeff2012}. 
The QPB was obtained using filter wheel closed event lists provided by the ESAS calibration database. For each region, background spectra were extracted for each EPIC camera at the same position in instrumental coordinates using the \textit{pn-spectra} and \textit{mos-spectra} tasks. These spectra were then normalized to the level of QPB in the observations, using \textit{pn\_back} and \textit{mos\_back}. Spectrum counts are grouped to have at least thirty counts per bin and then fitted using modified chi-square statistics. The errors are given by the confidence interval of the fits at 1\,$\sigma$.

In order to study the astrophysical background contributing to the measured flux we extracted spectra from several large regions surrounding the Arches cloud and fitted the corresponding spectra with a two-thermal-plasma model accounting for the Galactic thermal emission \citep[at 1 and 7\,keV, ][]{muno2004}. The ratio between the normalization of the soft and of the hot components has different values depending on the region we consider. This difference is to be linked to the strong anisotropy of the spatial distribution of the soft emission. Therefore, getting a precise estimate of the astrophysical background at the position of the Arches cloud is not straightforward. Nevertheless, using an elliptical region (referenced as `Bkg test' in Tab.\,\ref{tab:regions}) that has a sufficient coverage for all years, we fitted the normalization of the hot plasma (fixing the ratio between the two components at its mean value of 5.5). The weighted average of the hot plasma normalization is then $\rm I_{\rm 7keV} = 11.4\pm0.3 \times 10^{-4} \rm \, cm^{-5}$ with a maximal amplitude for the variation lower than 6.6\%. 

\begin{table}
        \centering
        \caption{Elliptical regions used for the spectral extraction.}
        \label{tab:regions}
        \begin{scriptsize}      
        \begin{tabular*}{0.45\textwidth}{@{\extracolsep{\fill}}c c c c c}
        \hline \hline
       Region & l [$^\circ$]& b [$^\circ$] & Axes [$''$] & Angle [$^\circ$] \\
        \hline
        Cloud & 0.124 & 0.018 & 58.9, 25.1 & 125.4 \\
        Cluster (excl.) & 0.123 & 0.019 & 16.0, 14.0 & 58.6 \\
        \hline
        Bkg test & 0.134 & --0.031 & 100.8, 48.3 & 125.42\\
        \hline
        NuStar & 0.122 & 0.019 & 50, 50 & -- \\
        \hline \hline
        \end{tabular*}
        \end{scriptsize}        
\end{table}

To account for the emission of the Arches cloud region we used the following model, 
\begin{equation}
	\centering
	\textsc{wabs}\times(\textsc{apec}+\textsc{powerlaw})+\textsc{gaussian}.
	\label{eq:mod1}
\end{equation}
A similar model was used by \cite{tatischeff2012} but here we do not correct the iron line emission for the overall absorption\footnote{The absorption correction relies on the \nh\ value fitted to the softer part of the spectrum, which also depends on the background subtraction. This would force the correlation between the line and the continuum emissions.}. According to the parameters derived for the Arches cloud from the overall \textit{XMM-Newton} data set  \citep[Table 3 in][]{tatischeff2012}, we fixed the temperature of the plasma component (\textsc{apec}) to kT~=~2.2~keV, the cloud metallicity to Z~=~1.7~Z$_\odot$, the index of the \textsc{powerlaw} to $\Gamma$~=~1.6, the centroid energy and width of the \textsc{gaussian} to E$_{\rm 6.4keV}$~=~6.4~keV and $\Delta\rm E_{\rm 6.4keV}$~=~10~eV, respectively. The free parameters are therefore the overall absorption and the normalizations of the plasma, of the line and of the reflection continuum components. For each year, we fit these four parameters between 2 and 7.5~keV on all spectra simultaneously. 
All fits were satisfactory, giving reduced $\chi^2 \sim 1$. We also tested a similar model including two thermal plasma at 1 and 7 keV, respectively, in order to better account for the astrophysical background present in the data. This second model gave consistent results regarding the \fe\ line emission. Interpreting the continuum component is more complex since it highly depends on the astrophysical background estimation. However, as discussed above, the steadiness of the astrophysical background is such that any variation larger than 10\% can only be attributed to a variation of the Arches cloud emission.  

\begin{figure}
	\centering
	\vspace{-0.4cm}
	\includegraphics[width=0.5\textwidth]{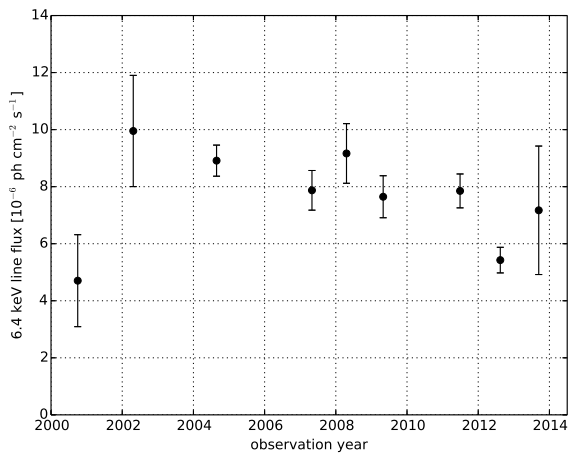}
	\caption{\fe\ line flux lightcurve of the Arches cloud. The emission is compatible with a constant emission up to 2011 with an average value of $\rm F_{\rm 6.4keV} = 8.2\times10^{-6}\rm \,ph\rm \,cm^{-2}\,\rm s^{-1}$ but the constant fit on the whole period is rejected at 4.3\,$\sigma$ due to a more than 30\% drop in 2012. 
	}
	\label{fig:FeKaLC}
\end{figure}

\begin{figure}
	\centering
	\vspace{-0.4cm}
	\includegraphics[width=0.5\textwidth]{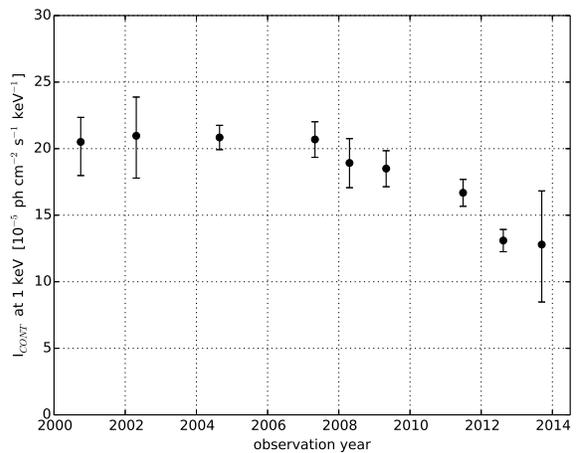}
	\caption{Continuum flux lightcurve associated with the power-law component of the Arches cloud. The lightcurve is compatible with a constant emission up to 2011 with an average value of $\rm I_{\rm cont} = 19.3\times10^{-5}\rm \,ph \rm \,cm^{-2}\rm \,s^{-1}\rm\,keV^{-1}$ but the constant fit on the whole period is rejected at 5.6\,$\sigma$.
	}
	\label{fig:ContLC}
\end{figure}

\section{Variation of the X-ray non-thermal emission}
\label{sec:analysisResults}
Fig.\,\ref{fig:ArchesFeKa} presents the \fe\ line emission of the Arches region for seven different years between 2000 and 2013 along with the best correlation found for the molecular material in the region. The overall X-ray emission is distributed within an elongated ellipse, centered on the Galactic east of the Arches cluster. This shape is fully consistent with the morphology of the \n2h\ (J=1--0) emission seen by Mopra around \hbox{--25~km~s$^{-1}$} \citep{jones2012}. However, its exact position is shifted by about 18$''$ compared to the molecular data. This is larger than the 10$''$ pointing error mentioned by \cite{jones2012}. Nevertheless, the higher antenna temperature in the Galactic south of the Arches ellipse is consistent with the brightest emission observed in the \fe\ line. This is in agreement with the fluxes reported by \cite{capelli2011b}, and with the position shift of the hard X-ray emission detected by \cite{krivonos2014} using \textit{NuSTAR}. We also point out that this south molecular core is not seen in the CS line at the corresponding velocity \citep{tsuboi1999}. This is a hint that the CS emission might be self absorbed at this position, indicating a rather dense core.

From Fig.\,\ref{fig:ArchesFeKa} it is also clear that the \fe\ line emission is varying over the 2000--2013 time period with both morphological and intensity changes. In particular, both the north and south subregions of the Arches cloud (dashed ellipses) seem to be increasing and then decreasing in flux, with a peak in 2004 and 2007, respectively. In 2012, there is an overall decrease of the emission. In spite of a relatively shallow exposure in 2013, the overall morphology looks quite different from what is seen in the previous years, since the emission seems to surround the \n2h\ dense cores. 

In order to confirm the variations seen in the images we performed spectral fits of the Arches cloud data (excluding the cluster) using the model detailed in Sec.\,\ref{sec:dataAnalysis}. \hbox{Figs.\,\ref{fig:FeKaLC} and \ref{fig:ContLC}} present the variations of the \fe\ line flux and of the power-law continuum emission, respectively.
Both components are varying significantly with a rejection of the constant fit at 4.3 and 5.6\,$\sigma$, respectively, and both have dropped by more than 30\% in 2012.
When a Pearson correlation test is applied to the power-law continuum flux as a function of the 6.4 keV flux (following the prescription of \citealt{pozzi2012} to take uncertainties into account), a correlation is detected at 3\,$\sigma$ confidence level and the linear fit results in a slope of $1.0\pm0.3$, indicating that most of the power-law continuum is indeed linked to the 6.4\,keV emission.
As expected in this case, the equivalent width of the \fe\ line is compatible with being constant over time with an average value ${\rm EW} = 0.9\pm0.1$\,keV
(rejection at less than 0.2\,$\sigma$).
The two other free parameters of the spectral fit are compatible with being constant over the thirteen-year period and the weighted mean values are $\rm N_{\rm H}=6.0\pm0.3\times10^{22}\rm\,cm^{-2}$ for the absorption and $\rm I_{\rm 2.2keV} = 3.6\pm0.7\times10^{-4}\rm\,cm^{-5}$ for the normalization of the thermal plasma. We point out that the normalization of the plasma given here does not consider the 2007 value that is significantly higher due to a contamination from the flaring cluster \citep{capelli2011a}.   

To compare our values to the ones presented in \cite{tatischeff2012}, we also divided the 2004 data set in two time periods and performed a constant fit of the data up to 2009. For this restricted data set, both the \fe\ line flux and the power-law continuum normalization are compatible with being constant (rejection at less than $1.4\,\sigma$) and result in $\rm F_{\rm 6.4keV}=8.3\pm1.0\times10^{-6}\rm\,ph\,cm^{-2}\,\rm s^{-1}$ ($9.6\pm1.0$, if corrected for the absorption) and $\rm I_{\rm cont}=20.1\pm1.6\times10^{-5}\rm\,ph\,cm^{-2}\,\rm s^{-1}\,\rm keV^{-1}$, respectively. Both values are compatible with \cite{tatischeff2012} results even if the continuum component cannot be directly compared due to different background estimations in the two analyses. 
The individual data points are also compatible within the error bars, except for the 2000 point for which values are about 2\,$\sigma$ apart. We can exclude that this discrepancy is due to a different position of the source within the field of view compared to the 2002--2009 observations. Indeed the systematic error associated\footnote{This systematic error is related to the effective area of the EPIC camera \citep{tatischeff2012}. We tested this effect on our data set by computing the flux of the Sgr~A East region as a function of its offset within the detector. We found a maximal amplitude $<$\,4\% for this systematic.} is negligible compared to the 1\,$\sigma$ error bars given by the fit.
The difference is most likely due to the poor quality of the corresponding data set but it has not been fully identified. In any case, we point out that excluding the 2000 data point does not change the significance of the constant fit rejection. 

The \textit{XMM-Newton} observations of autumn 2012 can also be compared to the \textit{NuSTAR} observations collected in the same period \citep{krivonos2014}. To do so, we extracted spectra from a circular region of 50$''$ radius centered on the cluster (referred as `NuStar' in Tab.\,\ref{tab:regions}) and fitted them with the same model (eq.\,\ref{eq:mod1}) but fixing the overall absorption, the temperature of the plasma and the index of the power-law to the \textit{NuSTAR} values \citep[Table 5 and Model 1 in][]{krivonos2014}. The values found for the \fe\ line flux (corrected for absorption, $\rm F_{\rm 6.4keV}=1.02\pm0.06\times10^{-5}\,\rm ph\, cm^{-2}\,\rm s^{-1}$), for the normalization of the plasma ($\rm I_{\rm 1.76keV}=50\pm1\times10^{-4}\,\rm cm^{-5}$) and of the continuum ($\rm I_{\rm cont}=1.58\pm0.08\times10^{-12}\,\rm erg\,cm^{-2}\,\rm s^{-1}$ over 3--20 keV) are fully compatible with the values stated by \cite{krivonos2014} for the same region. Since the absolute cross calibration factor between \textit{XMM-Newton} and \textit{NuSTAR} is less than 3\% \citep{risaliti2013}, we infer that the apparent consistency between \cite{tatischeff2012} and \cite{krivonos2014} fluxes is to be linked to the larger region considered in the later work. The \textit{NuSTAR} value is therefore consistent with the 2012 emission drop.

\section{Discussion: origin of the non-thermal emission}
\label{sec:discussion}
The non-thermal emission measured in the region surrounding the Arches cluster could be created by   either a particle bombardment or a strong hard X-ray irradiation by an external source. Since the spectrum of the Arches cloud does not provide enough information to discriminate between the two models \citep{capelli2011b,tatischeff2012,krivonos2014}, the remaining diagnostics are linked to the location of the emission and its variability.

The absence of correlation between this non-thermal X-ray emission and known molecular features at other wavelengths was supporting the particle bombardment scenario. In this work we identified for the first time a relevant molecular counterpart of the 6.4\,keV emission using the \n2h\ tracer. In particular, the overall morphology of the molecular structure with a two-lobe shape is in good agreement with the X-ray image. 
This supports the reflection scenario. The peak to peak correlation is not fully verified because of the slight shift mentioned in Section\,\ref{sec:analysisResults}. 
However, the exact distribution of the gas might be different than the one given by this molecular tracer, and the complex propagation of the X-ray signal within this structure may also account for the displacement \citep{clavel2013}.

Furthermore, the 30\% decrease within about one year, detected for both the 6.4 keV line and the continuum emissions, can only be explained by the reflection of hard X-ray irradiation. Indeed, the diffusion timescale and the possible rate of energy losses of the low-energy cosmic ray protons only allow for an emission decrease on much longer time scales (decades) and the energetic required for low-energy cosmic ray electrons is not realistic \citep{tatischeff2012}. Therefore, the significant variation of the non-thermal emission detected for the first time in the present work is the key element to conclude that a significant fraction of this emission is due to reflection. 

The source at the origin of the X-ray emission from the Arches cloud is unlikely to be located within the Arches cluster \citep{capelli2011b,tatischeff2012}. Moreover, the variability observed is not isolated since variations have already been observed in nearby structures \citep{capelli2011b} and, on larger scale, within the Sgr~A region \citep{ponti2010,clavel2013}. A large fraction of the non-thermal emission of the Arches cloud is therefore likely due to the past activity of \sgr, as is the emission of most of the regions listed above.

\section*{Acknowledgments}

The scientific results reported in this article are based on observations obtained with \textit{XMM-Newton}, an ESA science mission with instruments and contributions directly funded by
ESA Member States and NASA. The molecular map was obtained using the Mopra radio telescope, a part of the Australia Telescope National Facility which is funded by the Commonwealth of Australia for operation as a National Facility managed by CSIRO. M.C. acknowledges the Universit\'e Paris Sud 11 for financial support. S.S. acknowledges the Centre National d'Etudes Spatiales (CNES) for financial support. G.P. acknowledges support via an EU Marie Curie Intra-European fellowship under contract no. FP-PEOPLE-2012-IEF-331095. Partial support through the COST action MP0905 Black Holes in a Violent Universe is acknowledged.

{}
\appendix

\section[]{Observation selection}

The observations used in the analysis are listed in Tab.\,\ref{tab:obsID}, along with the corresponding clean exposure and the fraction of the Arches cloud region covered by each instrument. Any instrument presenting either less than 7 ks clean exposure or less than 80\% coverage has been discarded from the spectral analysis. We preferred to exclude observations with a low coverage in order to avoid a bias due to the non uniform spatial distribution of the emission within the region.

\begin{table} 
        \centering
        \caption{\textit{XMM-Newton} observations used in the present work.}
        \label{tab:obsID}
        \begin{scriptsize}
        \begin{tabular*}{0.49\textwidth}{c c c c c c c c}
        \hline \hline
       Date & Obs. ID & \multicolumn{3}{c}{Exposure$^{\rm a}$ [ks]} & \multicolumn{3}{c}{Coverage$^{\rm b}$} \\
       \cline{3-5} \cline{6-8}
          		&      		 &  m1 & m2 & pn & m1 & m2 & pn \\
        \hline
        2000-09-19 & 0112970401 & 22.2 & 22.1 & 18.6 & 1.00 & 1.00 & 0.90 \\
        2000-09-21 & 0112970501 & 14.0 & 14.6 &   5.3 & 0.97 & 0.99 & 0.91 \\
        \hline
        2002-02-26 & 0111350101 & 42.2 & 41.5 & 38.5 & 1.00 & 1.00 & 0.64 \\
        2002-10-03 & 0111350301 &   7.5 &   7.9 &   6.4 & 0.99 & 1.00 & 0.53 \\
        \hline
        2004-03-28 & 0202670501 & 32.9 & 30.4 & 14.8 	& 1.00 & 1.00 & 1.00 \\
        2004-03-30 & 0202670601 & 32.6 & 35.2 & 25.6 & 1.00 & 1.00 & 0.99 \\
        2004-08-31 & 0202670701 & 78.8 & 84.5 & 59.5 & 1.00 & 0.99 & 1.00 \\
        2004-09-02 & 0202670801 & 94.6 & 98.6 & 70.0 & 1.00 & 0.98 & 1.00 \\
        \hline
        2007-02-27 & 0506291201 & 22.7 & 24.8 & 0     & 1.00 & 0.90 & 0  \\
        2007-03-30 & 0402430701 & 26.3 & 28.0 & 18.1 & 0 & 1.00 & 1.00 \\
        2007-04-01 & 0402430301 & 60.7 & 62.9 & 53.3 & 0 & 1.00 & 0.99 \\
        2007-04-03 & 0402430401 & 40.6 & 41.2 & 26.7 & 0 & 1.00 & 1.00 \\
        2007-09-06 & 0504940201 &   8.9 &   9.3 &   5.9 & 1.00 & 1.00 & 0.39\\
        \hline
        2008-03-23 & 0505670101 & 73.5 & 74.3 & 49.0 & 0 & 1.00 & 1.00 \\
        \hline
        2009-04-01 & 0554750401 & 32.4 & 33.5 & 30.2 & 0 & 1.00 & 1.00 \\
        2009-04-03 & 0554750501 & 41.0 & 41.6 & 36.6 & 0 & 1.00 & 0.99 \\
        2009-04-05 & 0554750601 & 37.0 & 36.8 & 28.7 & 0 & 1.00 & 0.99 \\
        \hline
        2011-03-28 & 0604300601 & 31.3 & 32.7 & 24.1 & 0 & 1.00 & 0.99 \\
        2011-03-30 & 0604300701 & 37.3 & 41.9 & 22.6 & 0 & 1.00 & 1.00 \\
        2011-04-01 & 0604300801 & 35.1 & 34.9 & 31.5 & 0 & 1.00 & 1.00 \\
        2011-04-03 & 0604300901 & 21.4 & 22.4 & 15.3 & 0 & 1.00 & 0.99 \\
        2011-04-05 & 0604301001 & 39.8 & 41.6 & 23.6 & 0 & 1.00 & 0.99 \\
        2011-08-31 & 0658600101 & 47.6 & 48.0 & 45.2 & 1.00 & 1.00 & 0.45 \\
        2011-09-01 & 0658600201 & 40.4 & 43.0 & 37.0 & 1.00 & 1.00 & 0.47 \\
        \hline
        2012-03-13 & 0674600601 & 9.1 & 10.2 & 8.1 & 0 & 1.00 & 0.90 \\
        2012-03-15 & 0674600701 & 13.5 & 14.4 & 8.9 & 0 & 1.00 & 0.93 \\
        2012-03-17 & 0674601101 & 10.7 & 11.1 & 7.0 & 0 & 1.00 & 0.98 \\
        2012-03-19 & 0674600801 & 18.3 & 18.5 & 15.2 & 0 & 1.00 & 0.99 \\
        2012-03-21 & 0674601001 & 20.9 & 21.6 & 17.4 & 0 & 1.00 & 1.00 \\
        2012-08-31 & 0694640301 & 40.1 & 40.1 & 38.6 & 0 & 0.36 & 0.57 \\
        2012-09-02 & 0694640401 & 44.4 & 44.0 & 13.1 & 0.97 & 1.00 & 0.88 \\
        2012-09-24 & 0694641101 & 39.7 & 39.8 & 38.8 & 0.95 & 0.94 & 0.88 \\
        2012-09-26 & 0694641201 & 39.8 & 40.5 & 38.2 & 1.00 & 0.90 & 1.00 \\
        \hline
        2013-08-30 & 0724210201 & 42.6 & 43.7 & 40.5 & 0 & 1.00 & 0.50 \\
        2013-09-22 & 0724210501 & 32.7 & 32.9 & 26.5 & 0 & 1.00 & 0.48 \\
        \hline \hline
        \end{tabular*}
        \begin{flushleft}
		{\em $^{\rm a}$} Clean exposure at the Arches cloud position, for each of the EPIC camera instruments: MOS1 (m1), MOS2 (m2) and PN (pn).\\
		{\em $^{\rm b}$}  Fraction of the Arches cloud region covered by the observations.
		\end{flushleft}
        \end{scriptsize}        
\end{table}

\label{lastpage}


\begin{thebibliography}{}

\bibitem[\protect\citeauthoryear{Capelli et 
al.}{2011a}]{capelli2011a} Capelli R., Warwick R.~S., Cappelluti N., Gillessen S., Predehl P., Porquet D., Czesla S., 2011a, A\&A, 525, L2 

\bibitem[\protect\citeauthoryear{Capelli et 
al.}{2011b}]{capelli2011b} Capelli R., Warwick R.~S., Porquet D., Gillessen S., Predehl P., 2011b, A\&A, 530, A38 

\bibitem[\protect\citeauthoryear{Clavel et 
al.}{2013}]{clavel2013} Clavel M., Terrier R., Goldwurm A., Morris M.~R., Ponti G., Soldi S., Trap G., 2013, A\&A, 558, A32 

\bibitem[\protect\citeauthoryear{Jones et al.}{2012}]{jones2012} 
Jones P.~A., et al., 2012, MNRAS, 419, 2961 

\bibitem[\protect\citeauthoryear{Krivonos et 
al.}{2014}]{krivonos2014} Krivonos R.~A., et al., 2014, ApJ, 781, 
107 

\bibitem[\protect\citeauthoryear{Muno et al.}{2004}]{muno2004} 
Muno M.~P., et al., 2004, ApJ, 613, 326 

\bibitem[\protect\citeauthoryear{Ponti et al.}{2010}]{ponti2010} 
Ponti G., Terrier R., Goldwurm A., Belanger G., Trap G., 2010, ApJ, 714, 
732 

\bibitem[\protect\citeauthoryear{Ponti et al.}{2013}]{ponti2013} 
Ponti G., Morris M.~R., Terrier R., Goldwurm A., 2013, ASSP, 34, 331 

\bibitem[\protect\citeauthoryear{Pozzi, Di Matteo, 
\& Aste}{2012}]{pozzi2012} Pozzi F., Di Matteo T., Aste T., 2012, EPJB, 85, 175 

\bibitem[\protect\citeauthoryear{Risaliti et 
al.}{2013}]{risaliti2013} Risaliti G., et al., 2013, Nature, 494, 449 

\bibitem[\protect\citeauthoryear{Snowden et 
al.}{2008}]{snowden2008} Snowden S.~L., Mushotzky R.~F., Kuntz K.~D., Davis D.~S., 2008, A\&A, 478, 615 

\bibitem[\protect\citeauthoryear{Tatischeff, Decourchelle, 
\& Maurin}{2012}]{tatischeff2012} Tatischeff V., Decourchelle A., Maurin G., 2012, A\&A, 546, A88 

\bibitem[\protect\citeauthoryear{Tsuboi, Handa, 
\& Ukita}{1999}]{tsuboi1999} Tsuboi M., Handa T., Ukita N., 1999, ApJS, 120, 1 

\bibitem[\protect\citeauthoryear{Tsujimoto, Hyodo, 
\& Koyama}{2007}]{tsujimoto2007} Tsujimoto M., Hyodo Y., Koyama K., 2007, PASJ, 59, 229 

\bibitem[\protect\citeauthoryear{Wang, Dong, 
\& Lang}{2006}]{wang2006} Wang Q.~D., Dong H., Lang C., 2006, MNRAS, 371, 38 

\bibitem[\protect\citeauthoryear{Yusef-Zadeh et 
al.}{2002}]{yusef-zadeh2002b} Yusef-Zadeh F., Law C., Wardle M., Wang 
Q.~D., Fruscione A., Lang C.~C., Cotera A., 2002, ApJ, 570, 665 

\end{thebibliography}
\end{document}